# Photofield electron emission from an optical fiber nanotip

S. Keramati,[1,a)] A. Passian,[2,3,a)] V. Khullar,[2] and H. Batelaan[1,a)]

[1]*Department of Physics and Astronomy, University of Nebraska-Lincoln, Lincoln, Nebraska, 68588, USA*

[2]*Quantum Information Science Group, Oak Ridge National Laboratory, Oak Ridge, Tennessee, 37831, USA*

[3]*Department of Physics and Astronomy, University of Tennessee, Knoxville, Tennessee, 37996, USA*

**Abstract.** We demonstrate a nanotip electron source based on a graded index multimode silica optical fiber, tapered at one end to a radius of curvature r ~ 50 nm and coated with a thin film of gold. We report observation of laser-induced electron photoemission at tip bias potentials below the onset of dark field emission (FE). Single-photon photofield emission (PFE) is identified as the emission mechanism, which exhibits fast switching times with an upper limit on the order of 1 μs. The explored fiber optic nanotips are flexible back-illuminated emitters, which can be operated in continuous wave (CW) and pulsed modes using lasers with photon energies in the visible range or higher. The mechanical flexibility of the source can facilitate externally controlled positioning. Multiple, individually addressable, nanotips may be assembled into a bundle for applications such as computational electron ghost imaging.

Nanotip electron photocathodes have been widely studied in recent years as point-like emitters. This is incentivized by the advances in laser technology, which has enabled joint time-resolved and spatially coherent operation[1—5]. Commonly, a pulsed laser beam is sent into the vacuum chamber hosting the needle source, where it is tightly focused onto the nanotip apex[6—8]. This makes the beam alignment on the tip challenging and prone to mechanical vibrations. Grating-coupled metallic nanotips, which involve plasmonic nanofocusing, have been studied, partly to circumvent the requirement of direct apex illumination[9]. A few back-illuminated tip sources, for which some strict optical alignment requirements are circumvented, have been realized. For example, a 100-μm-long diamond nanotip needle, back-illuminated with a low-repetition-rate ns laser in the UV range, was shown to photoemit in the short-wavelength (i.e., single-photon absorption) regime[10]. Other works include the demonstration of an optical-fiber-based tungsten (W) (10 nm) nanotip source, in which the employed low-power continuous-wave (CW) diode laser triggered electron emission, and the significant difference in current rise time (~ 0.01 s) and the fast optical switching time is indicative of thermionic emission[11]. The reported emission was observed from a multimode fiber having a sub-wavelength aperture at its tapered apex, a system that is

---

[a)] Authors to whom correspondence should be addressed. Electronic mail: (S.K) sam.keramati@huskers.unl.edu; (A.P) passianan@ornl.gov; (H.B) hbatelaan@unl.edu.

ordinarily used as the optical probe of a scanning near-field optical microscope (SNOM). In a similar study using a large core optical fiber that had a gold (Au) coating on its flat end core surface, electron emission was observed following injection of fs short-wavelength UV laser pulses[12]. The studied emitter falls in the category of flat photocathodes illuminated from backside.

We recently investigated a tapered Au-coated fiber optic nanotip as an electron source. Using the 50-fs output pulses of an optical parametric amplifier (OPA) in the visible range, we achieved fast wavelength-dependent pulsed electron emission from the proposed nanotip[13]. From the analysis of the emission properties, we identified the photoemission mechanism to be surface plasmon resonance (SPR) enhanced above-threshold emission (ATE), through hot carrier saturation, and within a narrow (~ 20 nm) spectral range around the SPR design-wavelength. Multiphoton photoemission (MPE) was dominant across the rest of the spectrum. Unlike in the previous three examples, where comparatively large tip bias (extraction) potentials (order of 1 kV) were present, the field enhancement due to the SPR excitation rendered high electrostatic fields and large laser fluences unnecessary.

In the present work, we seek to study the electron emission properties from a similarly fabricated fiber tip, this time within the low-power regime (order of 0.1—1 mW) of CW lasers. SPR induced emission at such low intensities is not a major contributor. To compare, in the pulsed experiment[13] mentioned above, the laser intensity delivered to the tip during ATE excitation is estimated at $9 \times 10^{14}$ W/m$^2$, corresponding to an electric field of ~ 1 V/nm. In contrast, at the indicated range of CW input powers, the optical intensity is lower by 7 to 8 orders of magnitude. This is 2 to 3 orders of magnitude smaller than the smallest reported peak intensity in the literature which has given rise to continuous-wave multiphoton photoemission[14]. Here, we demonstrate photoemission in the CW regime through a single-photon photofield emission (PFE) process at photon energies which are smaller than the Au work function of ~ 5.1 eV[15]. Specifically, we will investigate emission driven by wavelength/energy = 405 nm/3.06 eV, 532 nm/2.33 eV, 633 nm/1.96 eV, and 672 nm/1.85 nm. We focused our experimental efforts on extracting a complete set of data at $\lambda$ = 633 nm line of a He-Ne laser. The remaining three (diode) lasers were observed to give rise to similar emission trends, albeit, under identical conditions, a larger photocurrent amplitude is observed at higher photon energies. This is expected in light of the considerable steepness of the Fermi-Dirac spectral carrier density distribution about the metallic Fermi level at room temperature.

The graded index multimode fiber (Corning, InfiniCor 600) with core/cladding diameter of (50/125 µm) and NA = 0.20 was tapered under laser heating using a micropipette puller (Sutter Instruments). The tapered region was coated with Cr (4 nm) followed by Au (100 nm) by electron vacuum evaporation. The Cr coating is used as an adhesion layer making the interface more robust. The experimental setup is shown

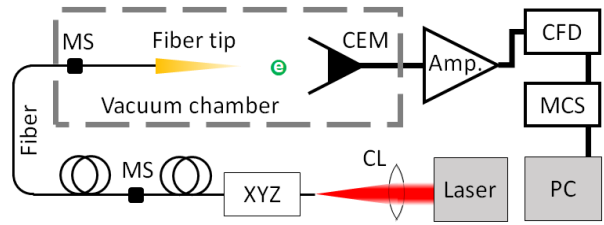

FIG. 1. Schematic representation of the experimental setup. Aided by a collimating lens (CL), a translation stage (XYZ), and mating sleeves (MS), the output beam of a laser is coupled into the fiber located inside the vacuum chamber, demarcated by the dashed line. The emitted photoelectrons (*e*) are detected by a channel electron multiplier (CEM), the output of which is routed through an amplifier (Amp.), a constant fraction discriminator (CFD), and a multichannel scaler (MCS).

schematically in Fig. 1.

The CW laser beam is lens-coupled into a strand of fiber. The coupling end of this fiber is inserted into a 250-μm ferrule of a fiber chuck mounted on a 3D (XYZ) micro-stage. All the reported power values in this manuscript were measured at its terminated end. An SMA-SMA mating sleeve (MS) is used to end-fire (i.e. no lens used) couple the terminated end of this fiber to a home-built fiber optic vacuum feedthrough, made using an identical type of fiber and sealed with vacuum epoxy. The experiments were performed at a pressure of ~ $2\times10^{-7}$ Torr. A high-vacuum-rated MS delivers the light to the fiber tip. A transmission of ~ 80% was estimated before the fiber tip. For a similar 30-cm long Au fiber tip (with the tapered shank of ~ 2-mm long), the transmission was ~ 0.3%. The electrons are detected in the pulse counting mode using a channel electron multiplier (CEM) (Dr. Sjuts, Model KBL 510) in front of the tip apex at a distance of 1 cm. The amplified signals are fed into a constant fraction discriminator (CFD). A multichannel scaler (MCS), triggered by the optical chopper TTL signals, records the CFD output logic pulses. A chopper wheel was mounted after the coupling lens (CL). The dark (i.e. DC) field emission (FE) was observed to start at a tip voltage of $V_{dark}$ ~ -450 V. The voltage was applied using a fine piece of copper wire attached by silver paint to a point located a few cm away from the apex.

The experimental results are displayed in Fig. 2. The power-dependence of the CW photoemission rate is shown in Fig. 2(a). When the laser light is blocked by a chopper blade, the count rate is approximately zero given that the electrostatic bias field is set to not trigger dark FE. The tip voltage $V_{tip}$-dependence of the photocurrent at a fixed input power (the maximum laser power in panel (a)) is shown in Fig. 2(b). A switching behavior is observed. As the triangular potential becomes progressively narrower by raising the magnitude of the bias field, the emission current increases accordingly. At $V_{tip}$ = -475 V, a dark current appears, as indicated by the measured counts in the absence of the laser light. The laser beam (~ 5 mm dia.) was focused onto the plane of the chopper wheel using a converging lens. The smaller cross-sectional area (~ 1 mm dia.) reduces the optical switching time. The estimated rise time of 8 μs in Fig. 2(c) is illustrated by the red arrows. This upper limit is consistent with the fast photoemission process. As photoemission is a single-photon process, it is linear in intensity, as confirmed by the power-law trend in Fig. 2(d). Each data point in this panel is the average of the corresponding 9 data points of the left-most peak (centered around t = 3 ms) in panel (a). To verify PFE is the dominant mechanism, we also inspected the voltage dependence (Fig. 2(e)). The blue circles correspond to the Fowler-Nordheim (FN) dark FE response of the tip (laser off). At each applied $V_{tip}$, 6 consecutive MCS bins (see Fig. 1) of duration 5 s were averaged. The errorbars correspond to one standard deviation. The red curve is the fitted FN model[15,16]. The FN count rate, for a tip radius $r$ and a hemispherical emission site with radius $R$, is given by

$$C_{FN} = \left(\frac{2\pi R^2}{q}\right) \times \left(\frac{aF^2}{\phi q^2 t^2}\right) \times \exp\left(\frac{-vbq\phi^{3/2}}{F}\right), \tag{1}$$

in which $q$ is the elementary charge, $\phi$ is the Au workfunction in the absence of applied static field (~ 5.1 eV), $a = q^3/(8\pi h)$ and $b = 8\pi\sqrt{2m}/(3qh)$, where $h$ is the Planck constant, and $F = qV_{tip}/kr$ is the force per unit charge on the tip, where $k = 5.7$ accounts for the typical geometrical shape of the metallic nanotips[6–8]. The parameters $t$ and $v$ are[15]

$$t = 1 + \frac{1}{9}y^2(1 - \ln y),$$
$$v = (1 - y^2) + \frac{1}{3}y^2 \ln y, \tag{2}$$

where $y = \sqrt{q^2 F/4\pi\varepsilon_0\phi^2}$ is the Nordheim parameter, and $\varepsilon_0$ is the vacuum permittivity. The fitted curve to the dark FE data (red line) in Fig. 2(e) is obtained for $r = 42.2\ nm$ and $R = 53.0\ nm$, as the only free parameters in this model. The extrapolated tip radius $r$ is indeed consistent with the scanning electron microscope (SEM) image of the tip shown in Fig. 3(a). The black-square data points in Fig.2 (e) are each the average of the 9 data points of the left-most peak (centered around t = 3 ms) in panel (b) for the 5 different values of $V_{tip}$. The PFE fit to this set of data is obtained as follows. An $n$-photon absorption process is equivalent to one in which the workfunction is reduced by $nh\nu$, followed by the corresponding power-law term. In principle, for the (long-wavelength) photon energy in the present case, 0- to 2-photon processes are allowed, and in general, the transition probability decreases by several orders of magnitude for larger values of $n$. The count rate in the PFE model is therefore expressed as

$$C_{PFE} = a_0 C_{FN,0} + a_1 C_{FN,1} I + a_2 C_{FN,2} I^2, \tag{3}$$

in which $C_{FN,n}$, $n = 0,1,2$ is the FN dark count rate of Eq. (1) with $\phi$ replaced with $\phi - nh\nu$. We set $a_0 = 1$, $a_1 = 5.89 \times 10^{-17}\ W^{-1}m^2$, and $a_2 = 10^{-4} \times (a_1)^2$, whereupon the satisfactory PFE fit (green line) in Fig. 2(e) is achieved. The intensity at the apex is estimated at $I = 5.40 \times 10^8\ Wm^{-2}$. With these parameters, the two-photon rate is negligible. The Schottky-reduced workfunctions in Fig. 2(e) fall between 3.4 eV and 3.6 eV. This is calculated from $\varphi = \phi - \sqrt{4QF}$, where $Q = \alpha\hbar c/4$ in which $\alpha$ is the fine structure constant, $\hbar$ is the reduced Planck constant, and $c$ is the speed of light in vacuum (hence $\alpha\hbar c = 1.44\ eV \cdot nm$ )[15].

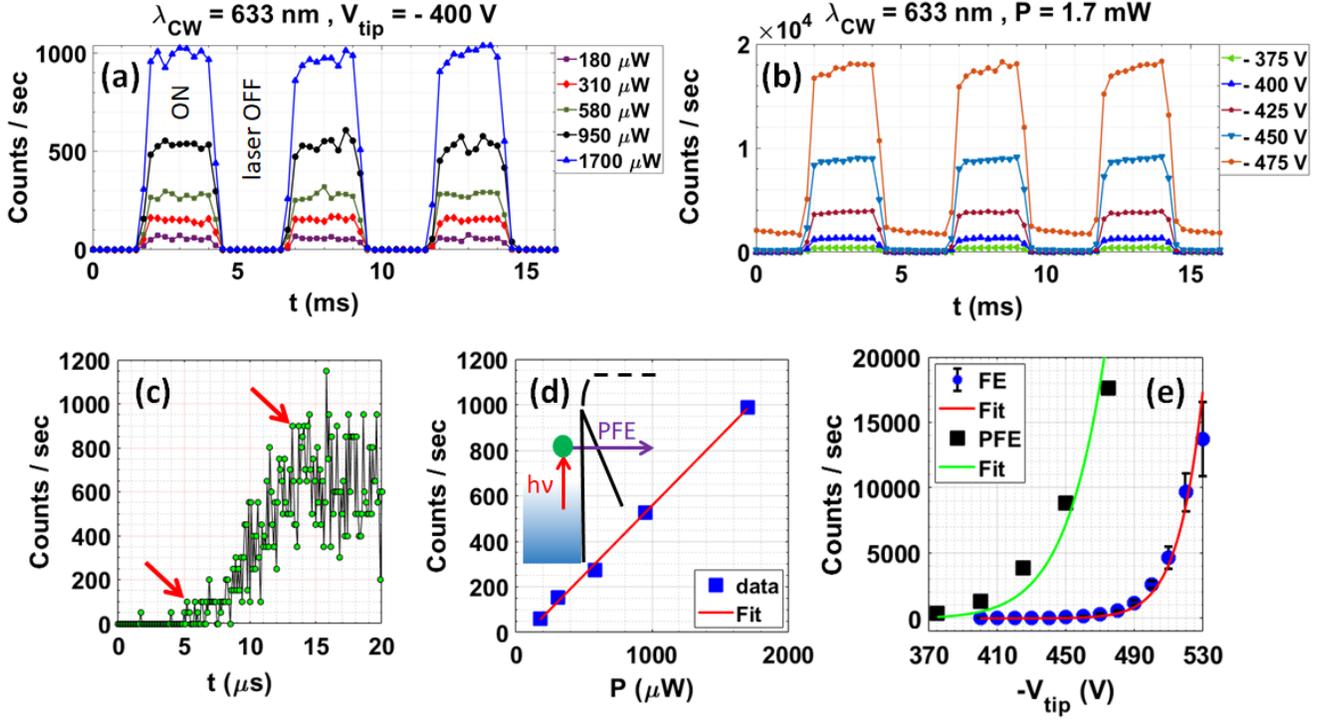

FIG. 2. Experimental data detailing the observed electron emission. Shown are: (a) switching at different input CW laser powers, (b) switching at different tip bias potentials, (c) fast switching with a rise time ~ 8 μs (estimated from the pair of data points singled out by the red arrows), (d) linear power-law trend indicating a single-photon-absorption process extracted from panel (a), and (e) the Fowler-Nordheim (FN) field emission (FE) along with the photofield emission (PFE) process extracted from panel (b). Blue circles display ±1σ errorbars, while black squares present the single-photon photofield emission (PFE), each with their corresponding fit model. The inset in (d) depicts the PFE process, which constitutes FN tunneling of an electron (green circle) across the Schottky-reduced potential preceded by absorption of a single photon of energy hν.

The SEM image of the tip taken before the experiments is shown in Fig. 3(a). During the course of the experiments, we intermittently check the $V_{tip}$ corresponding to the onset of dark FE which remained at $V_{dark}$ ~ -450V for this tip over several weeks. After performing the experiments with all the CW lasers, higher laser intensities were applied to explore any potential structural damage that may occur, as shown in Fig. 3(b). Damage is primarily heralded by a sudden abrupt jump in $V_{dark}$ to over -1 kV. The subsequently acquired SEM image shown in Fig. 3(b), confirmed the coating had been damaged. Several fiber tips exhibited similar damage pattern. SEM images show an undamaged tip when $V_{dark}$ remains the same, while damage is always observed after $V_{dark}$ jumps to a larger value. For a damaged tip such as the one in Fig. 3(b), no switching behavior with any appreciable contrast between the laser ON/OFF cycles, similar to Fig. 2(a,b), was observed.

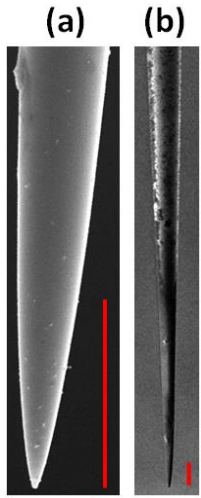

FIG. 3. Electron microscopy of the fiber tip (a) before the experiments, and (b) after long periods of experiment when the tip ultimately and abruptly got damaged at larger intensity exposures and bias potentials. Both scale bars = 10 μm. The uniform bright segment is the gold coating (left). While the tip shank was fully coated before the experiment, after the experiments, the gold coating near the tip is almost gone (right). Only discrete bright segments on the upper shank about 50-100 μm above the tip apex are visible. The image on the right was zoomed out to find these last remnants of gold coating.

The demonstrated fiber tip is flexible. In principle, it is possible to raster scan such a point-like electron source anywhere inside the vacuum chamber. A time-resolved nano-scale probe, that can deliver combinations of electron and light pulses is thus feasible. In addition, inexpensive CW diode lasers, widely available in the visible range, can be used to achieve CW photocurrents at modest applied bias potentials below the onset of dark FE. Another advantage of such fiber tips is the possibility to bundle them in an independently addressable matrix, a system of interest for computational ghost imaging applications[17], which was recently demonstrated for the first time for electrons[18]. Among the promises that ghost imaging offers are the replacement of imaging cameras with simple "bucket" detectors and to provide imaging with reduced exposure to the object. The latter is important for objects such as biological tissues that degrade under exposure. The recent proof-of-principle demonstration[18] was limited to millimeter spatial resolution. To be able to implement ghost imaging in electron microscopes, a switchable grid of electron sources must be imaged on the specimen with nanometer-scale spatial resolution. We envision that a fiber-bundle of nanotips, in which each fiber is individually coupled to a CW laser, may be used as the electron source. The possibility to implement CW laser beams modulated at arbitrary rates is particularly relevant here. Ghost imaging requires the illumination of all the pixels of an object by one electron for each image taken. As a result, say, for a modest, 10×10 image, about 100 electrons are required. Or, in other words, each fiber needs to emit $10^4$ cps for $10^4$ images per second to be captured. This corresponds to a repetition period of 100 μs, which is shown to be achievable in this work.

In summary, from the observed electron emission characteristics, we conclude that the introduced metallized fiber optic taper presents a versatile nanotip electron source. The flexibility afforded by the back-illumination, as opposed to external illumination, not only circumvents the difficulties arising from light scattering that may contaminate the measurements but also furnishes nanoscale tip positioning and maneuverability. From the presented analysis, we identify single-photon PFE as the dominant mechanism for the observed fast emission process. With further optimization, including more advanced design and fabrication procedures including possible nanostructuring of the tip as in nanoantenna formation, we envision, heretofore impossible or challenging, applications in scanning probe microscopy and spectroscopy (e.g. as in pump-probe measurement). An important feature of the explored nanotips is their configurability into a fiber bundle, where individual tips may be triggered independently, or groups of tips may be addressed collectively. Such features are of particular interest, for example, in computational electron ghost imaging applications.

**Data availability statement.** The data that supports the findings of this study are available from the corresponding authors upon reasonable request.

**Supplementary Material.** See Supplementary Material for further results.

**Acknowledgements.** A. Passian acknowledges support from the Laboratory Directed Research and Development Program at Oak Ridge National Laboratory (ORNL). ORNL is managed by UT- Battelle, LLC, for the US DOE under contract DE-AC05- 00OR22725. The fiber optic probes were fabricated at ORNL. S. Keramati and H. Batelaan acknowledge support by a UNL Collaborative Initiative grant, and by the National Science Foundation (NSF) under the award numbers EPS-1430519 and PHY-1912504. The SEM images were taken at the NanoEngineering Research Core Facility (NERCF), which is partially funded by the Nebraska Research Initiative. We thank Dr. P. Lougovski at ORNL for useful discussions. S. Keramati would like to thank Dr. C. Uiterwaal for helpful discussions.